\documentclass[12pt,a4paper,twoside]{article}

\newcommand{\be}{\begin{equation}}
\newcommand{\ee}{\end{equation}}
\newcommand{\Tr}{{\rm Tr}}

\newcommand{\Ah}{{\hat A}}
\newcommand{\Ab}{{\bar A}}

\newcommand{\JMP}{{\it J.~Math.~Phys.~}}
\newcommand{\CMP}{{\it Comm.~Math.~Phys.~}}
\newcommand{\JHEP}{{\it JHEP~}}
\newcommand{\NP}{{\it Nucl.~Phys.~}}

\newcommand{\MPL}{{\it Mod.~Phys.~Lett.~}}
\newcommand{\PR}{{\it Phys.~Rev.~}}
\expandafter\ifx\csname mathbbm\endcsname\relax

\else

\fi
\begin{document}
\begin{titlepage}
\begin{flushleft}  
       \hfill                      {\tt hep-th/0007043}\\
       \hfill                      UUITP-07/00\\
       \hfill                      July 2000\\
\end{flushleft}
\vspace*{3mm}
\begin{center}
{\LARGE Flux tube solutions in noncommutative gauge theories
\\}
\vspace*{12mm}
{\large Alexios P. Polychronakos\footnote{E-mail:
poly@teorfys.uu.se} \\
\vspace*{5mm}
{\em Theoretical Physics Department, Uppsala University \\
Box 803, S-751 08  Uppsala, Sweden \/}\\
\vspace*{5mm}
and \\
\vspace*{5mm}
{\em Physics Department, University of Ioannina \\
45110 Ioannina, Greece\/}\\}
\vspace*{15mm}
\end{center}

\begin{abstract}
We derive nonperturbative classical solutions of noncommutative $U(1)$ 
gauge theory, with or without a Higgs field, representing static magnetic 
flux tubes with arbitrary cross-section. The fields are nonperturbatively
different from the vacuum in at least some region of space. The flux of 
these tubes is quantized in natural units.
\end{abstract}

\end{titlepage}

Noncommutative gauge theories are the topic of much recent
interest \cite{CDS}-\cite{IIKK}. Early examples of such theories arose in
nonperturbative regularizations of membranes \cite{WHN,FFZ}, D-branes and 
strings \cite{W}-\cite{L}, and they have also recently emerged at various
limits of M theory. They are, therefore, relevant to aspects of string 
and D-brane dynamics. They exhibit an interesting space uncertainty 
relation \cite{Y} and have many similarities with large-$N$ gauge 
theories and lattice gauge theories \cite{BM,MRS}.

The above theories are field theories on a noncommutative space 
\cite{C} (for a concise review see also \cite{D}). For a flat such 
space the coordinates $X_i$ obey the commutation relations
\be
[ X_i , X_j ] = i \theta_{ij}
\ee
with $\theta_{ij}$ a set of commuting parameters. Specializing to
3+1 dimensions, and assuming a commutative time dimension, we can
always bring the above to the form
\be
[X,Y]=i\theta ~,~~~[X,Z] = [Y,Z] =0
\ee
with $\theta>0$. So only the $(X,Y)$ plane is noncommutative and 
resembles a quantum
phase space (with $\theta$ playing the role of $\hbar$) on which
$X$ and $Y$ act as coordinate and momentum operators. 

One expects that at the limit $\theta \to
0$ such theories go over to ordinary commutative ones. There can be,
however, interesting nonperturbative effects that arise for nonzero
$\theta$ and do not have a smooth limit. These effects could be present
already at the classical level, in the form of nonperturbative solutions.
Several such solutions have recently appeared \cite{NS}-\cite{GN}. 
In \cite{GN}, in particular, a solution representing a magnetic monopole 
has been derived. In this letter we will demonstrate the existence of 
nonperturbative
solutions corresponding to magnetic flux tubes of arbitrary cross-section.

One way to describe field theories on noncommutative spaces is through
star products of ordinary functions. In this approach, a one-to-one 
correspondence
between functions of operators $\hat f$ and ordinary functions $f$
is established by Weyl ordering:
\be
{\hat f} (X,Y) = \frac{1}{2\pi} \int d^2 k {\tilde f} (k_1 , k_2 )
e^{i(k_1 X + k_2 Y)}
\ee
where $\tilde f $ is the Fourier transform of $f$. The product of two
Weyl ordered operators ${\hat f}{\hat g}$, Weyl reordered, corresponds
to a function denoted $f * g$ 
\be
f * g \, (x,y) = \left. e^{\frac{i}{2}\theta (\partial_{x_1} 
\partial_{y_2} - \partial_{x_2} \partial_{y_1})} f(x_1 , y_1 ) 
g(x_2 , y_2 ) \right|_{x_1 = x_2 , y_1 = y_2}
\ee
which defines the noncommutative star product. 
One then writes noncommutative field theory lagrangians in terms of 
commutative fields but with products replaced by star products.

We will specialize to the case of noncommutative $U(1)$ gauge theories,
with or without Higgs fields. The action involves the gauge fields
$A_\mu$ and, possibly, a Higgs field $\Phi$:
\be
S = \frac{1}{4g^2} \int d^4 x \, \left\{ F_{\mu \nu}^2 
+ ( D_\mu \Phi )^2 \right\}
\ee
where
\be
F_{\mu \nu} = \partial_\mu A_\nu -\partial_\nu A_\mu + i ( A_\mu * A_\nu
- A_\nu * A_\mu )
\ee
and 
\be
D_\mu \Phi = \partial_\mu \Phi + i (A_\mu * \Phi - \Phi * A_\mu )
\ee
Indices $(0,1,2,3) = (t,x,y,z)$ are summed with the flat Minkowski metric.
(We used the fact $\int dx dy \, f * g = \int dx dy \, fg$ to eliminate
star products in $S$.)
In the standard commutative limit the above is a free theory, $g$ is
irrelevant and the
Higgs field $\Phi$ decouples from the Maxwell field. Here, however, we
have a coupled interacting system.

An alternative approach to noncommutative theories is to acknowledge that
they refer to operator-valued fields and work directly with operators
on the quantum phase space $(X,Y)$ which are functions of $z$ and $t$
\cite{AMN,GN,AGW}.
In this approach, star products become operator products and integration
over the $(X,Y)$ plane becomes trace:
\be
\int dx dy \, f(x,y) = 2\pi \theta \, \Tr {\hat f} (X,Y)
\ee
Derivatives in the $X$ and $Y$ direction
become the standard quantum generators of translations in these variables
and they act on operators in the adjoint way, via commutation.
That is, for any operator $F$
\be 
\partial_x F = \frac{i}{\theta} [Y,F] ~,~~~
\partial_y F = -\frac{i}{\theta} [X,F] ~,~~~
\ee
Understanding, now, $A_\mu$ and $\Phi$ as anti-hermitian operator-valued
functions of $z$ and $t$, the action becomes
\be
S = -\frac{\pi \theta}{4g^2} \int dz dt \, \Tr \left\{ F_{\mu \nu}^2 
+ ( D_\mu \Phi )^2 \right\}
\label{Sop}
\ee
with
\begin{eqnarray}
F_{\mu \nu} &=& \partial_\mu A_\nu - \partial_\nu A_\mu + [A_\mu , A_\nu ]
\\
D_\mu \Phi &=& \partial_\mu \Phi + {[} A_\mu , \Phi {]}
\end{eqnarray}
We can also write the field strength, in analogy to the commutative
case, as the commutator of affine derivative operators:
\be
{\hat F}_{\mu \nu} = [\Ah_\mu , \Ah_\nu ] 
\ee
where we defined
\be
\Ah_\mu = \partial_\mu + A_\mu
\label{Af}
\ee
We should stress that partial derivatives $\partial_X$ and $\partial_Y$,
when they act in the adjoint on operator fields commute. Viewed, however,
as operators on the same footing as $A_x$ and $A_y$, as we assume in 
(\ref{Af}), do {\it not} commute; $\hat F$, then, differs from $F$ by the
commutator of the two derivatives on the noncommutative plane:
\be
{\hat F}_{\mu \nu} = F_{\mu \nu} + \omega_{\mu \nu}
\ee
$\omega$ is a constant magnetic field in the $z$-direction
(proportional to the symplectic 2-form on the $(X,Y)$ plane) 
with one `quantum' of flux per unit cell of area $2\pi \theta$:
\be
\omega_{\mu \nu} = -\frac{i}{\theta} (\delta_{\mu 1} \delta_{\nu 2}
- \delta_{\nu 1} \delta_{\mu 2} )
\ee
Covariant derivatives also become
\be
D_\mu \Phi = [\Ah_\mu , \Phi ] 
\ee
This rewriting may be of little interest for the $z$- (and $t$-) components,
since $\partial_z$ and $A_z$ are operators on different spaces, but it is
quite relevant for the $X$ and $Y$ components since both $\partial_i$
and $A_i$ are operators on the same space; $\Ah_i$ are the only
operators appearing in the theory.

The action in terms of the new field strength becomes
\begin{eqnarray}
{\hat S} &=& -\frac{\pi\theta}{2g^2} \int dz dt ~\Tr \left\{ [\Ah_\mu , 
\Ah_\nu ]^2 + [\Ah_\mu , \Phi ]^2 \right\}
\nonumber \\
&=& S + \frac{\pi}{g^2 \theta} \int dz dt ~[ \Tr 1 - i \theta F_{12} ]
\end{eqnarray}
The first additional term is an irrelevant (infinite) constant while
the second is the trace of a commutator which contributes only ``boundary''
terms. The two actions lead to the same equations of motion:
\begin{eqnarray}
[\Ah_\nu , [\Ah_\nu , \Ah_\mu ]] + [\Phi , [\Phi , \Ah_\mu ]] &=& 0
\nonumber \\
{[}\Ah_\nu , {[}\Ah_\nu , \Phi {]}{]} &=& 0
\label{eom}
\end{eqnarray}
For static, purely magnetic configurations ($A_0 = 0$, $\partial_0 = 0$)
we can also write the Bogomolny equations
\be
[\Ah_i , \Ah_j ] = \pm \epsilon_{ijk} [\Ah_k , \Phi ] 
\label{Bog}
\ee
which minimize the energy of such configurations and automatically
satisfy the equations of motion (\ref{eom}) by virtue of the Jacobi identity.

We will look for solutions of (\ref{eom},\ref{Bog}) that represent static
magnetic flux tubes in the $z$-direction. We therefore take 
$\partial_z A_i = 0$ and also make the gauge choice $A_z = 0$. 
Defining the new fields
\be
A = \frac{1}{\sqrt 2} (\Ah_1 + i \Ah_2 ) ~,~~~
\Ab = \frac{1}{\sqrt 2} (\Ah_1 - i \Ah_2 )
\ee
the equations of motion for pure gauge theory simply become
\be
[A , [\Ab , A]] = 0
\label{AAA}
\ee
while the Bogomolny equations for the Higgs theory become
\be
0=\partial_z A = \pm [A, \Phi] ~,~~~ \partial_z \Phi = \pm [A, \Ab ]
\label{Bogz}
\ee
The second equation in (\ref{Bogz}) is solved as 
\be
\Phi = \pm z [A, \Ab ] + \Phi_0
\ee
with $\Phi_0$ a constant operator. The first equation, then,
becomes 
\be
[A , [\Ab , A]] = 0 ~,~~~ [A,\Phi_0 ]=[\Ab , \Phi_0 ] = 0
\ee
Thus the equation for $A$ is the same as in the theory without Higgs,
and we have an extra $\Phi_0$ field (which can be proportional to the
unit operator and does not contribute to the action or energy).

Instead of working with $X$ and $Y$ it is more convenient to work with 
oscillator operators
\be
a= \frac{X+iY}{\sqrt{2\theta}} ~,~~
a^\dagger = \frac{X-iY}{\sqrt{2\theta}} ~;~~~ [a,a^\dagger ] = 1
\ee
acting on oscillator states $|n>$, $n=0,1,\dots$ diagonalizing the number
operator $N = a^\dagger a$. In terms of these, $A$ and $\Ab$ become
\be
A = \frac{a}{\sqrt\theta} + A_1 + i A_2 ~,~~~
\Ab = \frac{a^\dagger}{\sqrt\theta} + A_1 - i A_2 
\ee
Note that $N$ plays the role of the square of the distance from the
origin $R$, $N \sim \frac{R^2}{2\theta}$.

The `vacuum' configuration $A_i = 0$ corresponds to $A = a/\sqrt\theta$. 
We shall look for rotationally invariant solutions in the form 
of a deformed oscillator \cite{P} 
\be
A = f(N) \, a
\ee
with $f(\cdot)$ a scalar function. The above is a shorthand for
the explicit expression
\be
A = \sum_{n=0}^\infty g(n) |n><n+1| ~,~~~ g(n) = \sqrt{n+1} f(n)
\ee
Substituting the above ansatz in (\ref{AAA}) we obtain for $g(n)$:
\be
g(n) \left\{ |g(n+1)|^2 - 2 |g(n)|^2 + |g(n-1)|^2 \right\} = 0
\ee
valid for all $n$ and subject to the boundary condition $g(-1) = 0$.
Thus, either $g(n)$ or the expression in brackets (which is the discrete
laplacian on $g$) should vanish. 

Assume that two successive zeros of $g(n)$ are $n_1 < n_2$. 
Then the above equation for $n_1 < n < n_2$ gives a system of 
$n_2 - n_1 -1$ linear homogeneous independent equations for the quantities 
$|g(n_1 +1)|^2 , \dots |g(n_2 -1)|^2$, which implies that all these $g(n)$
have to vanish as well. We conclude that the general situation is 
$g(-1) = \dots = g(n_0 -1 ) = 0$ with $n_0$ some nonnegative
integer. The rest of the $g(n)$ are essentially the same as those
for the operator $a$, only shifted up by $n_0$. Choosing vanishing phases
for $g(n)$ and an overall normalization $1/\sqrt\theta$ we obtain
\be
g(n) = \sqrt\frac{n-n_0 +1}{\theta} ~,~~~ n \geq n_0 ~;~~~ g(n)=0
~,~~~ 0 \leq n < n_0
\ee
$A$ can be expressed as
\be
A = \frac{a}{\sqrt\theta} \, \sqrt{\frac{N}{N-n_0}} \, {\hat {\rm P}}_{n_0}
\label{Asol}
\ee
where 
\be
\hat {\rm P}_{n_0} = 1- {\rm P}_{n_0} = \sum_{n=n_0}^\infty |n><n|
\ee
is the operator that projects out the first $n_0$ levels. Clearly
$n_0 = 0$ reproduces the vacuum solution $A=a/\sqrt\theta$. 
For large distances from the origin, that is, for $N\gg1$, $A$ becomes
\be
A \simeq \frac{a}{\sqrt\theta} \left(1+\frac{n_0}{2N} \right)
\sim \frac{a}{\sqrt\theta} + \frac{n_0}{\sqrt 2} \frac{X+iY}{R^2}
\label{Alim}
\ee
so it reproduces the gauge field of a Dirac string of strength $n_0$.
(Remember that $A_i$ are anti-hermitian, so (\ref{Alim}) implies that
$iA_j \sim \epsilon_{jk} X_k /R^2$, a vortex configuration.)

The field strength $\hat F$ is
\be
{\hat F}_{\mu \nu} = \omega_{\mu \nu} {\hat {\rm P}}_{n_0}
\ee
and so 
\be
F_{\mu \nu} = -\omega_{\mu \nu} {\rm P}_{n_0}
\ee
The above represents a circular static magnetic flux tube centered
at the origin of the $(X,Y)$ plane with radius $\sim \sqrt{\theta n_0}$.
It resembles a circular quantum Hall `droplet' with $n_0$ particles.
The total magnetic flux is
\be
\phi = 2\pi\theta \, \Tr (i F_{12}) = -2\pi n_0
\ee
and is obviously quantized. We stress that the obtained solution is
nonperturbative since, in the area around the origin, it cannot be
written as $A=a/{\sqrt\theta} + {\cal O} (\theta^0 )$.

This solution is by no means unique. Firstly, we can perform
any unitary transformation and still have a solution. Such deformations
are considered to be generalized gauge transformations, but they lead
to configurations with varying profiles on the $(X,Y)$ plane; for instance,
the coherent-squeezed state operator
\be
U = e^{\mu^* a - \mu a^\dagger} e^{\lambda^* a^2 - \lambda {a^\dagger}^2}
\label{U}
\ee
deforms the cross-section of the tube into an ellipsoidal one and
also moves it to a different position on the plane. In general, 
conjugation with all possible $U$ will change the position and shape
of the cross-section into any closed curve with the same area (the set
of all $U$ generates all area-preserving diffeomorphisms on the 
noncommutative space).
Thus all quantum Hall droplets correspond to flux tube solutions.

(Note that $U$ in (\ref{U}) above alters the asymptotic behavior
of $A$ for $N\gg 1$ and thus would lead to a field nonperturbatively
different from the vacuum everywhere. We can, however, choose
appropriate operators $U'$ which perform the required reshaping of the
tube while leaving the asymptotics of $A$ unchanged. Such a $U'$
would be, e.g.,
\be
U' = U^{f(N)}
\ee
with $U$ as above and $f(N)$ a real function such that $f(N) \to 0$ for 
$N \to \infty$ while $f(N) = 1$ for $N \sim n_0$.)

Secondly, we can start with similar higher-moment ansatze, such as
\be
A = f(N) \, a^k = \sum_n g(n) |n><n+k|
\ee
The analysis is similar, but now the equations group into $k$
uncoupled subsystems, each involving the coefficients $g(kn+q)$
for a single $q=0,1,\dots k-1$. We can choose a solution as above
for each $q$, that is,
\be
g(kn+q) = C_q \sqrt{n-n_q +1} ~,~~~ n \geq n_q
\ee
with different normalizations $C_q$ and shifts $n_q$ for each $q$.
The field strength $\hat F$ of such configurations becomes
\be
{\hat F}_{\mu \nu} = \theta \omega_{\mu \nu} \sum_{q=0}^{k-1} |C_q |^2
\sum_{n=n_q}^\infty |kn+q><kn+q|
\ee
So by choosing all $|C_q |^2 = 1/\theta$ we obtain a localized $F$
\be
F_{\mu \nu} = -\omega_{\mu \nu} \sum_{q=0}^{k-1} \sum_{n=0}^{n_q -1}
|kn+q><kn+q|
\ee
This is a flux tube with spatial extent $\sim \sqrt {\theta k n_{\rm max}}$,
where $n_{\rm max}$ is the maximum of $n_q$, and a varying radial profile. 
(We can, e.g., obtain an annular profile between the values $n_1$ and $n_2$
by choosing $k=n_2$ and $n_q = 0$ for $q<n_1$, $n_q = 1$ for $n_1 \le q< n_2$.)
We note, however, that the asymptotics of $A$ are always different than the
vacuum, so this seems to be an entirely nonperturbative solution (although
we have not excluded the possibility that an appropriate unitary
transformation will make it asymptotically perturbative).

Finally, we can take advantage of the scaling symmetry of the equations
of motion (\ref{eom}) when expressed in terms of $\Ah_i$ and simply
multiply the previous solutions by any number. Taking the original
circular-profile solution (\ref{Asol}) and multiplying by $\lambda$
gives a solution with a field strength
\be
F_\lambda = (\lambda^2 -1) \omega - \lambda^2 \omega {\rm P}_{n_0}
= -\omega {\rm P}_{n_0} + (\lambda^2 -1) \omega {\hat {\rm P}}_{n_0}
\ee
It represents a flux tube identical to the original one plus a constant 
magnetic field outside it. By choosing $\lambda^2 = 1 + {\cal O}
(\theta)$ we can make the constant field perturbative, if we wish.
We point out that this scaling transformation exists for any general
solution of the equations of motion, as
\be
\Ah_{x,y} (z,t) \to \lambda \Ah_{x,y} (\lambda z, \lambda t ) ~,~~~
A_{z,t} (z,t) \to \lambda A_{z,t} (\lambda z, \lambda t ) ~,~~~
\Phi (z,t) \to \lambda \Phi (\lambda z, \lambda t )
\ee
which leads to a transformation of the field strength
\be
F_{\mu \nu} (z,t) \to \lambda^2 F_{\mu \nu} (\lambda z, \lambda t ) 
+ (\lambda^2 -1) \omega_{\mu \nu}
\ee
We stress that this is a nonperturbative transformation which does
not go over to the usual scaling transformation of the commutative theory
at the limit $\theta \to 0$. Applied, for example, to the monopole
solution found in \cite{GN}, it produces a monopole with a {\it fractional}
strength $\lambda^2$ dilated in the $z$-direction by 
$\lambda^{-1}$, in the presence of a constant magnetic field.

In conclusion, we have explored several nonperturbative solutions of
the static equations of motion of noncommutative $U(1)$ gauge theory
with or without a Higgs field, in the shape of flux tube configurations.
The flux of these tubes seems to obey a quantization condition similar
to the one in compact electrodynamics, for which we have no topological
explanation. This is presumably related to the integrality of the index 
of some operator. The situation is certainly rather different from the
commutative case, since, for instance, we can evade the quantization of
the magnetic monopole charge by embedding it in a constant magnetic
field, as demonstrated in the previous paragraph. 

The analysis of this paper may also shed some light to the question 
of the stability of the monopole solution of \cite{GN}. The monopole
was tied to a string in the $z$-direction, asymptotically similar to 
our flux tube solution for $n_0 =1$. This string carries energy per
unit length, and would naturally want to contract and pull the monopole
in the $z$-direction. As we stressed, however, noncommutative gauge
theory is most naturally formulated in terms of $\Ah_i$ rather than $A_i$
(there is no natural separation of $\Ah_i$ into a $\partial_i$ and a
$A_i$ piece in the action). The field strength $\hat F$ in terms of
$\Ah$ has an additional constant magnetic field $\omega$. In the presence
of this field the string carries {\it no} field inside it and represents a 
{\it depletion} (rather than addition) of energy. It therefore has a
negative tension and wants to expand, pushing the monopole in the
negative $z$-direction. On the other hand, due to the magnetic field 
$\omega$, the monopole feels a magnetostatic force pulling it upwards.
This exactly balances the push of the string and stabilizes it.

Clearly the above heuristic discussion touches upon the question of
field self-interaction and charges in noncommutative gauge theories.
These and similar issues are left for future investigation.

\vskip 0.5cm
I would like to thank Ulf Lindstr\"om for interesting discussions.

\end{document}